\documentclass[prl,10pt,showpacs,amsmath,twocolumn,floatfix]{revtex4}

\usepackage{amsmath}           
\usepackage{amsfonts}          
\usepackage{amssymb}           
\usepackage{graphicx}          

\def\bs#1{\boldsymbol{#1}}

\begin{document}

\title{Quantum Discord and the Geometry of {Bell}-Diagonal States}

\author{Matthias D.~Lang}

\email{mlang@unm.edu}

\affiliation{Center for Quantum Information and Control, University
of New Mexico, MSC07--4220, Albuquerque, New Mexico 87131-0001, USA}

\author{Carlton M.~Caves}

\affiliation{Center for Quantum Information and Control, University
of New Mexico, MSC07--4220, Albuquerque, New Mexico 87131-0001, USA}

\date{\today}

\begin{abstract}
The set of Bell-diagonal states for two qubits can be depicted
as a tetrahedron in three dimensions.   We consider the level
surfaces of entanglement and quantum discord for Bell-diagonal
states.  This provides a complete picture of the structure of
entanglement and discord for this simple case and, in
particular, of their nonanalytic behavior under decoherence.
The pictorial approach also indicates how to show that discord
is neither convex nor concave.
\end{abstract}

\maketitle

Maintenance of quantum coherence is clearly important for
quantum-information-processing protocols.  Noise and
decoherence, by turning pure states into mixed states, generally
destroy quantum coherence.  Efficient representation of quantum
information requires that a quantum-information-processing
system be composed of parts~\cite{BlumeKohout}.  For
multi-partite systems, quantum coherence is related to
nonclassical correlations between the parts.

One kind of nonclassical correlation is
entanglement~\cite{entanglement}.  A pure quantum state is
unentangled if it is a product of pure states for each part.  A
mixed state is unentangled (separable) if it can be written as
an ensemble of such product states.  Entanglement is the crucial
resource for such quantum-information-processing protocols as
quantum key distribution, teleportation, and super-dense
coding~\cite{entanglement}.

Operational measures of entanglement are notoriously difficult
to calculate for mixed states; even the boundary between
separability and entanglement is difficult to characterize. One
can say, however, that the set of separable states is a convex
set, is invariant under local unitary operations, and has
dimension as large as the space of mixed
states~\cite{entanglement}.

Separable states have nonzero measure in the space of all
states~\cite{Zyczkowski}.  In a decoherence process that
involves decay to a separable equilibrium state that does not
lie on the boundary between separability and entanglement, the
decohering state will cross that boundary before reaching the
equilibrium state.  This phenomenon, dubbed ``sudden death of
entanglement''~\cite{esd,cole}, is the generic expectation in
view of the geometry of separable states.

Separable states can have nonclassical correlations even though
they are unentangled.  A state with only classical correlations,
often called a classical state, is one that is diagonal in a
product basis, for then the correlations are described by a
joint probability distribution for classical variables of the
parts.  These purely classical states are a set of measure zero,
as is suggested by the fact that any classical state can be
perturbed infinitesimally to become nonclassical by making two
of the eigenvectors infinitesimally entangled and is proved
rigorously in~\cite{Ferraro}.

A variety of measures have been proposed to quantify
nonclassical correlations for bipartite
systems~\cite{discord,measures,Modi}, in ways that can be
nonzero for separable, but nonclassical states. Nonclassical,
but perhaps separable correlations have been related to
exponential speed-ups in the ``power-of-one-qubit''
model~\cite{Knill} of mixed-state quantum
computation~\cite{Datta}, but the relation remains
tenuous~\cite{Dakic}.

One can use decoherence mechanisms to explore the nooks and
crannies of nonclassical-correlation measures.  There is no
sudden death~\cite{Ferraro}, as is suggested by the absence of
open sets of classical states, but the nonanalyticity of
nonclassical measures points to the possibility of sudden
changes in derivatives.  Investigation of the behavior of
nonclassical measures under decoherence has
begun~\cite{cole,Maziero2009,Maziero2010,Mazzola}, with a focus
on the action of decoherence within the class of two-qubit
states that are diagonal in the Bell basis.  This focus is
motivated by the fact that entanglement measures and
nonclassical-correlation measures can be calculated explicitly
for the Bell-diagonal states, thus allowing one to determine how
these measures change under decoherence.

The Bell-diagonal states are a three-parameter set, whose
geometry, including the subsets of separable and classical
subsets, can be depicted in three
dimensions~\cite{entanglement,RHorodecki}. Level surfaces of
entanglement and nonclassical measures can be plotted directly
on this three-dimensional geometry.  The result is a complete
picture, for this simple case, of the structure of entanglement
and nonclassicality.  We suggest that it is more illuminating to
use this picture to explain how measures of entanglement and
nonclassicality change along the one-dimensional trajectories
traced out by decohering states, rather than the other way
around. Hence we review and expand the pictorial approach here.

\begin{figure}[htbp]
\begin{center}
\includegraphics[width=6.25cm]{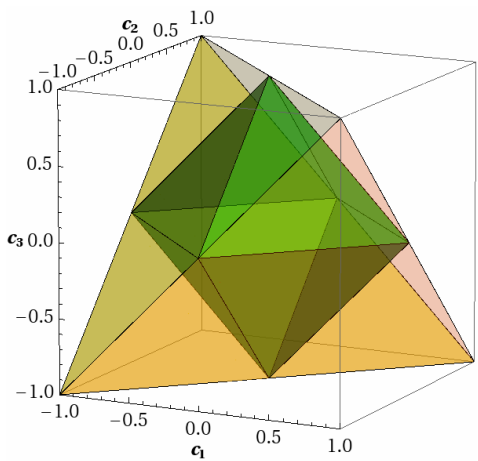}
\end{center}
\caption{Geometry of Bell-diagonal states.  The tetrahedron
${\cal T}$ is the set of valid Bell-diagonal states.  The Bell
states $|\beta_{ab}\rangle$ sit at the four vertices, the
extreme points of ${\cal T}$.  The green octahedron ${\cal O}$,
specified by $|c_1|+|c_2|+|c_3|\le1$ ($\lambda_{ab}\le1/2$), is
the set of separable Bell-diagonal states. There are four
entangled regions outside ${\cal O}$, one for each vertex of
$\cal T$, in each of which the biggest eigenvalue $\lambda_{ab}$
is the one associated with the Bell state at the vertex.
Classical states, i.e., those diagonal in a product basis, lie
on the Cartesian axes.\label{tetra}}
\end{figure}

The Bell-diagonal states of two qubits, $A$ and $B$, have density
operators of the form
\begin{align}
\rho_{AB}\!=\!
\frac{1}{4}\biggl(I+\sum^3_{j=1}c_j\,\sigma_j^A\otimes\sigma_j^B\biggr)
\!=\!\sum_{a,b}\lambda_{ab}|\beta_{ab}\rangle\langle\beta_{ab}|\;,
\label{states}
\end{align}
where the $\sigma_j$'s are Pauli operators.  The eigenstates are
the four Bell states
$|\beta_{ab}\rangle\equiv(|0,b\rangle+(-1)^a|1,1\oplus
b\rangle)/\sqrt2$, with eigenvalues
\begin{equation}
\lambda_{ab}=\frac{1}{4}\Bigl(1+(-1)^ac_1-(-1)^{a+b}c_2+(-1)^bc_3\Bigr)\;.
\end{equation}
Any two-qubit state satisfying
$\langle\sigma_j^A\rangle=0=\langle\sigma_j^B\rangle$, i.e.,
having maximally mixed marginal density operators
$\rho_A=I/2=\rho_B$, can be brought to Bell-diagonal form by
using local unitary operations on the two qubits to diagonalize
the correlation matrix
$\langle\sigma_j^A\otimes\sigma_k^B\rangle$.

A Bell-diagonal state is specified by a 3-tuple $(c_1,c_2,c_3)$.
The density operator $\rho_{AB}$ must be a positive operator,
i.e., $\lambda_{ab}\ge0$; the resulting region of Bell-diagonal
states is the state tetrahedron~${\cal T}$ in Fig.~\ref{tetra}.
Separable Bell-diagonal states are those with positive partial
transpose~\cite{entanglement}.  Partial transposition changes
the sign of $c_2$, so operators with positive partial transpose
occupy the reflection of $\cal T$ through the plane $c_2=0$; the
region of separable Bell-diagonal states is the intersection of
the two tetrahedra, which is the octahedron $\cal O$ of
Fig.~\ref{tetra}~\cite{RHorodecki}.

The entanglement of formation ${\cal
E}$~\cite{entanglement,Wootters} is a monotonically increasing
function of Wootters's concurrence~$C$~\cite{Wootters}, which
for Bell-diagonal states, is given by $C=\max(0,2\lambda_{\rm
max}-1)$, where $\lambda_{\rm max}=\max\lambda_{ab}$.  The
concurrence and the entanglement of formation are convex
functions on $\cal T$.  They are zero for the separable states
in the octahedron ${\cal O}$.  In each of the four entangled
regions outside ${\cal O}$, $C$ and $\cal E$ are constant on
planes parallel to the bounding face of ${\cal O}$ and increase
as one moves outward through these planes toward the Bell-state
vertex.

Quantum discord was introduced by Ollivier and
Zurek~\cite{discord}. We restrict attention to it because of its
prominence among measures of nonclassical correlations and
because it has been a focus of recent work on decoherence and
nonclassical
correlations~\cite{cole,Maziero2009,Maziero2010,Mazzola}.

To define quantum discord, one starts with the quantum mutual
information, ${\cal{I}}=S(\rho_A) + S(\rho_B) -
S(\rho_{AB})=S(\rho_B)-S(B|A)$, where $S(\rho) = -{\rm
tr}(\rho\log_2\rho)$ is the von Neumann entropy of $\rho$ and
$S(B|A)=S(\rho_{AB})-S(\rho_A)$ is a conditional quantum
entropy. The quantum mutual information is regarded as
quantifying the {\it total\/} correlations in the joint state
$\rho_{AB}$.

The quantum mutual information of Bell-diagonal states,
\begin{equation}
{\cal I}=2-S(\rho_{AB})
=\sum_{a,b}\lambda_{ab}\log_2(4\lambda_{ab})\;,
\end{equation}
is a convex function on $\cal T$.  It has smooth level surfaces
that bulge outward toward the vertices of $\cal T$.

The next step is to quantify {\it purely classical\/}
correlations in terms of information from measurements.  One
imagines measuring on~$A$ a POVM consisting of rank-one POVM
elements $E_k=Dq_k|k\rangle\langle k|$~\cite{Dattadiss}, where
$D$ is the dimension of $A$ and the $q_k$ make up a normalized
probability distribution.  The probability to get result $k$ is
$p_k=Dq_k\langle k|\rho_A|k\rangle$, and the post-measurement
state of $B$ is $\rho_{B|k}=\langle k|\rho_{AB}|k\rangle/\langle
k|\rho_A|k\rangle$.  Minimizing the average entropy of $B$,
given result $k$, over measurements on~$A$, yields a classical
conditional entropy
\begin{equation}
\tilde S(B|A)\equiv\min_{\{E_k\}}\sum_kp_kS(\rho_{B|k})\;;
\label{classcondent}
\end{equation}
minimizing chooses the measurement of $A$ that extracts as much
information as possible about~$B$.  The corresponding
mutual-information-like quantity ${\cal{C}}=S(\rho_B)-\tilde
S(B|A)$ is the measure of classical correlations.

For Bell-diagonal states, we have
\begin{equation}
{\cal C}\!=\!1-H_2\biggl(\frac{1+c}{2}\biggr)
\!=\!\frac{1+c}{2}\log_2(1+c)+\frac{1-c}{2}\log_2(1-c)\;,
\end{equation}
where $H_2(p)=-p\log_2p-(1-p)\log_2(1-p)$ is the binary entropy
and $c=\max|c_j|$~\cite{Luo,proof}.  This $\cal C$, a convex
function on $\cal T$, is constant on the surfaces of cubes (or
the portion of such a cube in $\cal T$) centered at the
origin of Fig.~\ref{tetra}---this introduces
nonanalyticity---and $\cal C$ increases monotonically with the
size of the cube.

Discord is defined as the difference of $\cal I$ and $\cal C$,
\begin{equation}
{\cal D}={\cal I}-{\cal C}=\tilde S(B|A)-S(B|A)\;,
\end{equation}
thus capturing a notion of nonclassical correlations.  Since
${\cal C}$ is generally asymmetric between $A$ and $B$, so also
is the discord; this means, in particular, that discord, as
defined, vanishes if and only if $\rho_{AB}$ is diagonal in a
conditional product basis
$|e_j^A\rangle\otimes|f_{jk}^B\rangle$, rather than only in a
product basis $|e_j^A\rangle\otimes|f_k^B\rangle$. Bell-diagonal
states being symmetric between $A$ and $B$, however, discord is
zero only for classical states, which lie on the Cartesian axes
in Fig.~\ref{tetra}~\cite{Dakic}.

\begin{figure}[htbp]
\begin{center}
\raisebox{17em}{(a)}\hspace{0em}\includegraphics[width=6.25cm]{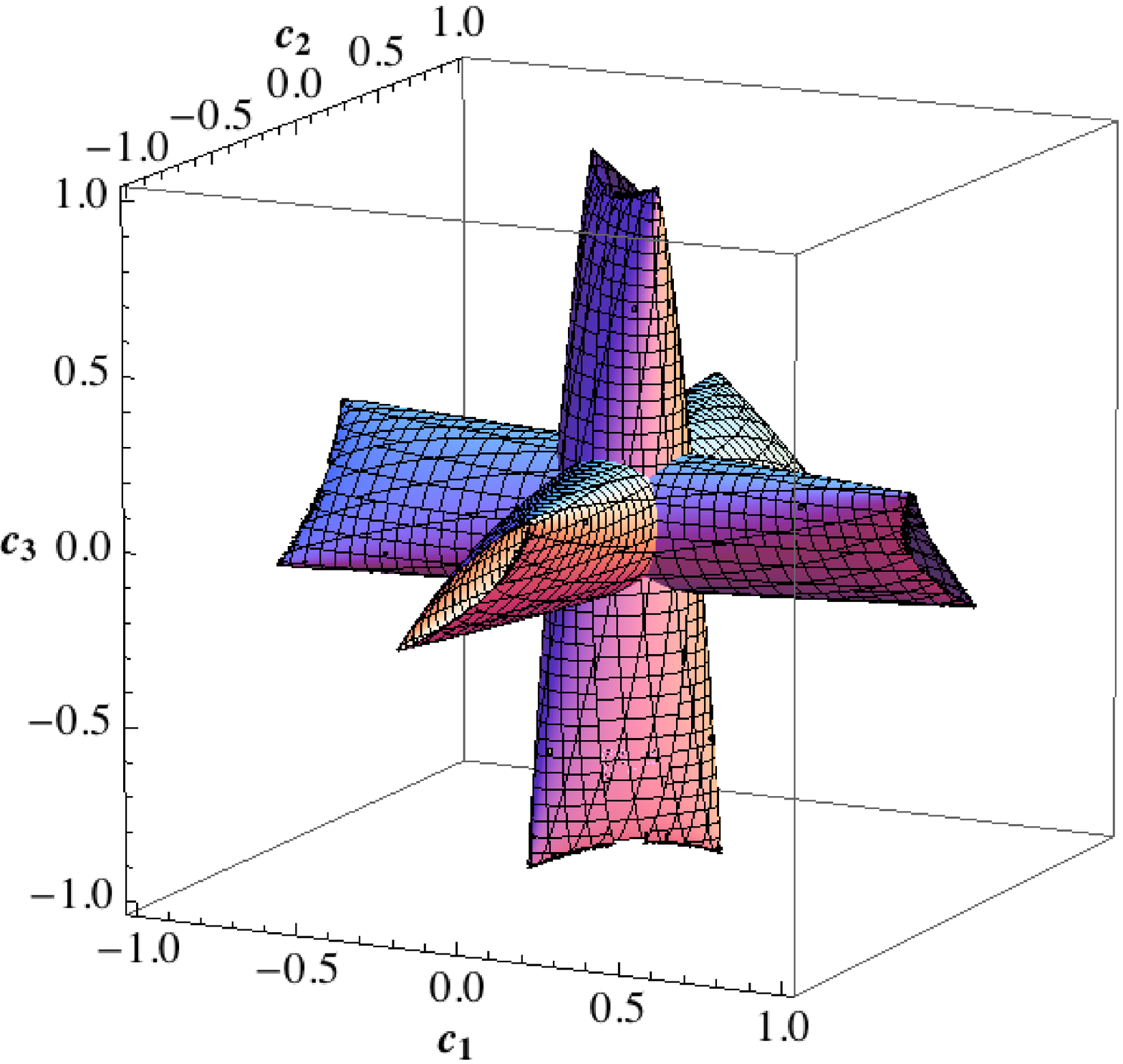}\\ 
\protect{\vspace{1em}}
\raisebox{17em}{(b)}\hspace{0em}\includegraphics[width=6.25cm]{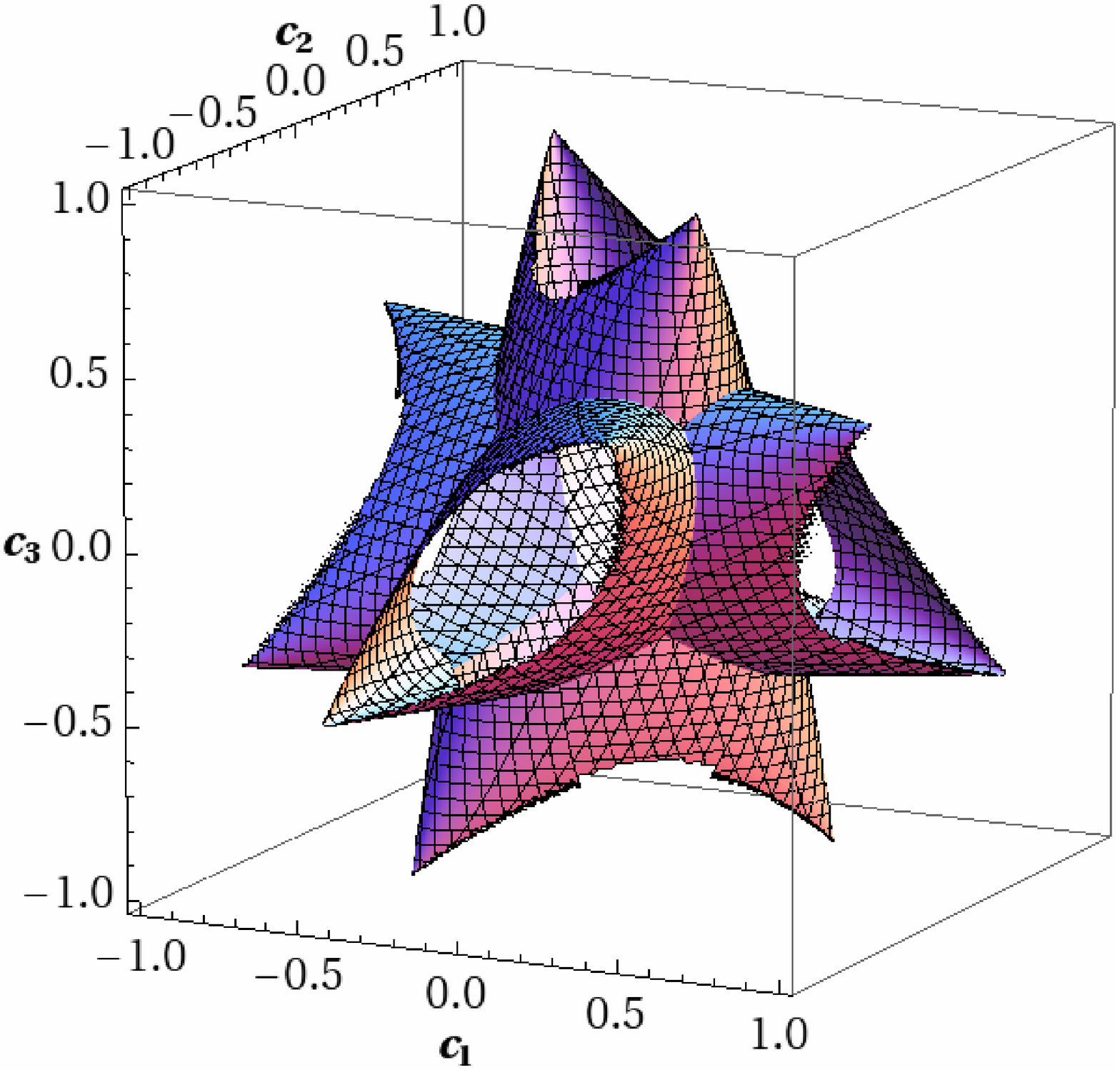}\\
\protect{\vspace{1em}}
\raisebox{17em}{(c)}\hspace{0em}\includegraphics[width=6.25cm]{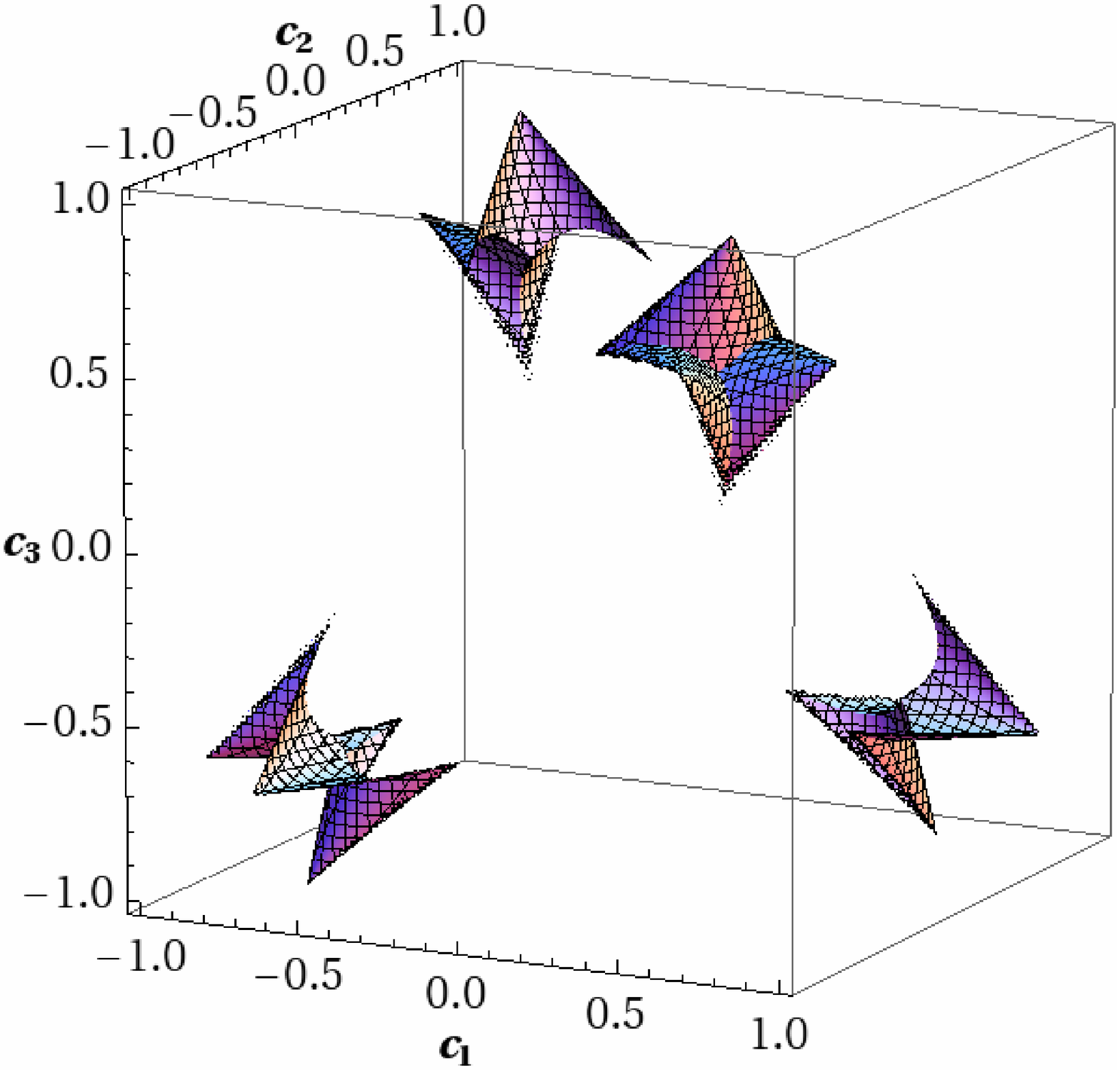}
\end{center}
\caption{Surfaces of constant discord: (a)~${\cal D}=0.03$;
(b)~${\cal D}=0.15$; (c)~${\cal D}=0.35$.  The level surfaces
consist of three intersecting ``tubes'' running along the three
Cartesian axes.  The tubes are cut off by the state tetrahedron
$\cal T$ at their ends, and they are squeezed and twisted so
that at their ends, they align with an edge of $\cal T$.  As
discord decreases, the tubes collapse to the Cartesian
axes~\protect\cite{Dakic}.  As discord increases, the tube
structure is obscured, as in~(c): the main body of each tube is
cut off by $\cal T$; all that remains are the tips, which reach
out toward the Bell-state vertices.\label{3surfaces}}
\end{figure}

Figure~\ref{3surfaces} plots level surfaces of discord for
Bell-diagonal states.  From these plots, it is clear that
discord is quite a different beast from entanglement of
formation, quantum mutual information, and the measure of
classical correlations. Whereas ${\cal E}$, ${\cal I}$, and
${\cal C}$ generally increase outward from the origin, ${\cal
D}$ increases away from the Cartesian axes, capturing an
entropic notion of distance from classical
states~\cite{Modi,Dakic}.  In particular, as one moves outward
along one of the constant-discord tubes of Fig.~\ref{3surfaces},
the classical correlations and the total correlations of the
quantum mutual information increase, but their difference, the
nonclassical correlations as measured by discord, remains
constant.  At the vertices of $\cal O$, ${\cal I}={\cal C}=1$
and ${\cal D}={\cal E}=0$.  At the Bell-state vertices of $\cal
T$, ${\cal I}=2$ and ${\cal C}={\cal D}={\cal E}=1$, this being
the maximum value of discord for two qubits.  In addition, $\cal
E$, $\cal I$, and $\cal C$ are all convex, whereas discord is
neither concave nor convex, as is evident from the plots in
Fig.~\ref{3surfaces}: one can mix two positive-discord states to
get a zero-discord classical state, and one can mix two
zero-discord classical states on different axes to get a
positive-discord state~\cite{note}.

\begin{figure}[htbp]
\begin{center}
\includegraphics[width=5.75cm]{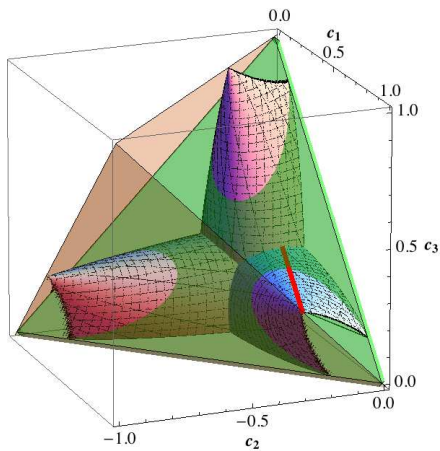}
\end{center}
\caption{Trajectory (red) of a Bell-diagonal state under random
phase flips of the two qubits; initial conditions are
$c_1(0)=1$, $-c_2(0)=c_3(0)=0.3$.  The trajectory is the
straight line $c_3=c_3(0)=0.3=-c_2/c_1$.  For clarity, only the
$(+,-,+)$-octant is shown.  A constant-discord surface is
plotted for the discord value of the initial state.  Faces of
the yellow state tetrahedron $\cal T$ and the green separable
octahedron $\cal O$ are also shown. The straight-line trajectory
proceeds along a tube of constant discord till it encounters the
vertical tube at $c_1=0.3$, after which discord decreases
monotonically to zero when the trajectory reaches the $c_3$
axis.  Entanglement of formation decreases monotonically to zero
when the trajectory enters $\cal O$ at $c_1=0.7/1.3=0.54$.
\label{trajectory}}
\end{figure}

Mazzola, Piilo, and Maniscalco~\cite{Mazzola} recently
investigated the dynamics of classical and nonclassical
correlations, as measured by discord, for two qubits under
decoherence processes that preserve Bell-diagonal states.  In
particular, they considered independent phase-flip channels for
the two qubits.  The phase flips are implemented mathematically
by random applications of $\sigma_z$ operators to the qubits.
This decoherence process leaves $c_3$ unchanged, but flips the
signs of $c_1$ and $c_2$ randomly, leading to exponential decay
of $c_1$ and $c_2$ at the same rate.  Mazzola and collaborators
found that for the initial conditions they considered, the
entanglement of formation decays to zero in a finite
time---sudden death of entanglement~\cite{esd}---but that the
discord remains constant for a finite time and then decays,
reaching zero at infinite time.  This situation is depicted in
terms of the surfaces of constant discord in
Figure~\ref{trajectory}.  The decohering-state trajectory is a
straight line that runs along a tube of constant discord, until
it encounters an intersecting tube, after which the discord
decreases to zero when the state becomes fully classical.

This behavior is generic for flip channels and initial
conditions on edges of the state tetrahedron.  We focus here on
the phase-flip channel with initial conditions in the
$(+,-,+)$-octant, but analogous considerations apply to the
other flip channels (bit and bit-phase) and to initial
conditions on the other edges of $\cal T$. Consider then initial
conditions anywhere along the edge of $\cal T$ in this octant:
$c_1(0)=1$ and $0\le-c_2(0)=c_3(0)\le1$.  The trajectory under
phase flips is a straight line $c_3=c_3(0)=-c_2/c_1$.  Along
this straight line, the eigenvalues $\lambda_{ab}$ factor into
products of probabilities, $(1\pm c_1)/2$ and $(1\pm c_3)/2$,
thus making $S(\rho_{AB})$ the entropy of two independent binary
random variables with these probabilities.  This yields a
quantum mutual information ${\cal
I}=2-H_2[(1+c_3)/2]-H_2[(1+c_1)/2]$. Furthermore, along the
trajectory $c=\max(c_1,c_3)$.  The result is that the trajectory
initially runs along a tube of constant discord
\begin{equation}
{\cal D}=1-H_2\biggl(\frac{1+c_3}{2}\biggr)\;,
\end{equation}
for $c_1\ge c_3$.  When $c_1=c_3$, the trajectory encounters
another tube, after which, for $c_1\le c_3$, the discord
decreases monotonically as ${\cal D}=1-H_2[(1+c_1)/2]$ as $c_1$
decreases. Meanwhile, the entanglement of formation decreases
monotonically from its initial value to a sudden death at
$c_1=(1-c_3)/(1+c_3)$.

The situation investigated in~\cite{Mazzola} is surely
interesting: under decoherence, nonclassical correlations remain
constant for a finite time interval.  This situation is,
however, a special one, as can be seen from the surfaces of
constant discord; the trajectories considered here are the only
straight lines in parameter space that stay on a surface of
constant discord.  Indeed, the pictorial approach can provide a
complete understanding of how entanglement and nonclassicality
change under decoherence within the set of Bell-diagonal states.

This work was supported in part by NSF Grant Nos.~PHY-0653596
and PHY-0903953.


\begin{thebibliography}{}

\bibitem{BlumeKohout}
R.~Blume-Kohout, C.~M. Caves, and I.~H. Deutsch, Found. Phys. {\bf
32}, 1641 (2002).

\bibitem{entanglement}
R.~Horodecki, P.~Horodecki, M.~Horodecki, and K.~Horodecki, Rev.
Mod. Phys. {\bf 81}, 865 (2009), and references therein.

\bibitem{Zyczkowski}
K.~{\.Z}yczkowski, P.~Horodecki, A.~Sanpera, and M.~Lewenstein, Phys.
Rev.~A {\bf 58}, 883 (1998).

\bibitem{esd}
T.~Yu and J.~H. Eberly, Science {\bf 323}, 598 (2009), and references
therein.

\bibitem{cole}
J.~H. Cole, J.~Phys.~A {\bf 43}, 135301 (2010).

\bibitem{Ferraro}
A.~Ferraro, L.~Aolita, D.~Cavalcanti, F.~M. Cucchietti, and
A.~Ac{\'\i}n, Phys. Rev.~A {\bf 81}, 052318 (2010).

\bibitem{discord}
H.~Ollivier and W.~H. Zurek, Phys. Rev. Lett \textbf{88}, 017901
(2001).

\bibitem{measures}
J.~Oppenheim, M.~Horodecki, P.~Horodecki, and R.~Horodecki,
Phys. Rev. Lett. {\bf 89}, 180402 (2002); S.~Luo, Phys. Rev.~A
\textbf{77}, 022301 (2008); M.~Piani, M.~Christandl, C.~E. Mora,
and P.~Horodecki, Phys. Rev. Lett. {\bf 102}, 250503 (2009);
S.~Wu, U.~V. Poulsen, and K.~M{\o}lmer, Phys. Rev.~A
\textbf{80}, 032319 (2009); A.~Brodutch and D.~R.~Terno, Phys.
Rev.~A {\bf 81}, 062103 (2010); M.~D. Lang, A.~Shaji, and C.~M.
Caves, ``Measures of nonclassical correlations,'' in
preparation.

\bibitem{Modi}
K.~Modi, T.~Paterek, W.~Son, V.~Vedral, M.~Williamson, Phys. Rev.
Lett. {\bf 104}, 080501 (2010).

\bibitem{Knill}
E.~Knill and R.~Laflamme, Phys. Rev. Lett. {\bf 81}, 5672 (1998).

\bibitem{Datta}
A.~Datta, A.~Shaji, and C.~M. Caves, Phys. Rev. Lett \textbf{100},
050502 (2008).

\bibitem{Dakic}
B.~Dak{\'\i}c, V.~Vedral, and {\v C}.~Brukner, {\tt arXiv:1004.0190
[quant-ph]}.

\bibitem{Maziero2009}
J.~Maziero, L.~C. C{\'e}leri, R.~M. Serra, and V.~Vedral,  Phys.
Rev.~A\/ {\bf 80}, 044102 (2009).

\bibitem{Maziero2010}
J.~Maziero, T.~Werlang, F.~F. Fanchini, L.~C. C{\'e}leri, and R.~M.
Serra, Phys. Rev.~A\/ {\bf 81}, 022116 (2010).

\bibitem{Mazzola}
L.~Mazzola, J.~Piilo, and S.~Maniscalco, Phys. Rev. Lett. {\bf
104}, 200401 (2010).

\bibitem{RHorodecki}
R.~Horodecki and M.~Horodecki, Phys. Rev.~A {\bf 54}, 1838 (1996).

\bibitem{Wootters}
W.~K.~Wootters, Phys. Rev. Lett. {\bf 80}, 2245 (1998).

\bibitem{Dattadiss}
One can begin with an arbitrary POVM, but the minimum in
Eq.~(\protect\ref{classcondent}) can always be attained on
rank-one POVM elements, as is shown in A.~Datta, PhD
dissertation, University of New Mexico (2008), {\tt
arXiv:0807.4490 [quant-ph]}.

\bibitem{Luo}
S.~Luo, Phys. Rev. A \textbf{77}, 042303 (2008).

\bibitem{proof}
Heretofore~\protect\cite{Luo}, ${\cal C}$ has been calculated
for Bell-diagonal states using only orthogonal projectors.  We
extend to rank-one POVMs here.  With POVM elements
$E_k=q_k(I+\bs{n}_k\cdot\bs{\sigma})$, we have $p_k=q_k$ and
$\rho_{B|k}=(I+\bs{d_k}\cdot\bs{\sigma})/2$, where
$d_{kj}=c_jn_{kj}$.  We have
$S(\rho_{B|k})=H_2[(1+|\bs{d_k}|)/2\ge H_2[(1+c)/2]$, since
$|\bs d_k|\le c$.  This shows that $\tilde S(B|A)\ge
H_2[(1+c)/2]$, with equality for measurement of orthogonal
projectors along the direction of maximum $c_j$.

\bibitem{note}
This argument and its conclusion are not special to
Bell-diagonal states: mixing two discordant states can lead to a
state that is diagonal in a product basis, and mixing two states
that are diagonal in incompatible product bases generally leads
to a discordant state.

\end{thebibliography}
\end{document}